\documentclass[twocolumn,aps,prb]{revtex4}
\usepackage{epsf,graphicx}

\begin{document}
\draft

\title{Finite size effects on thermal denaturation of globular proteins}
\author{Mai Suan Li$^1$, D. K. Klimov$^2$ and D. Thirumalai$^2$}

\address{$^1$Institute of Physics, Polish Academy of Sciences,
Al. Lotnikow 32/46, 02-668 Warsaw, Poland\\
$^2$Department of Chemistry and Biochemistry and Institute for Physical
Science and Technology, University of Maryland, College Park, MD 20742 }

\begin{abstract}

Finite size effects on the cooperative thermal denaturation of proteins
are considered. A dimensionless measure of cooperativity, $\Omega_c$, 
scales as $N^\zeta$, where $N$ is the number of amino acids. Surprisingly,
we find that $\zeta$ is universal with $\zeta = 1 + \gamma$, where the
exponent $\gamma$ characterizes the divergence of the susceptibility for a
self-avoiding walk. Our lattice model simulations and 
experimental data are consistent with the theory. Our finding rationalizes
the marginal stability of proteins and substantiates the earlier predictions
that the efficient folding of two-state proteins requires $T_F\approx
T_\theta$, where $T_\theta$ and $T_F$ are the collapse and folding
transition temperatures, respectively.

\end{abstract}

\maketitle



Single domain globular proteins, which are finite sized systems, undergo
remarkably cooperative transitions from an ensemble of unfolded states to
well ordered folded (or native) states as the temperature is lowered (Fig.
1(a)).  In many cases, the transition to the native state takes place in
an apparent two-state manner, i.e., the only detectable species are the
native (more precisely, the conformations belonging to the
native basin of attraction ({\bf NBA})) or unfolded ({\bf U}) states
\cite{Poland}.  Although the microscopic origin of cooperativity is
not fully understood \cite{Bryngelson95Proteins}, the transition to the
{\bf NBA} at the folding transition temperature, $T_F$, 
is a consequence of the effective interresidue attraction that
compensates for the entropy loss.  From this
perspective the {\bf NBA} $\leftrightarrow$ {\bf U} transition can be
viewed as a phase transition in a finite-sized system. 
Furthermore, the transition to the {\bf NBA} at $T_F$
has the characteristics of a first order phase transition
\cite{Poland,Bryngelson95Proteins}.  Many experiments have shown that
folded states of globular proteins are only marginally stable below $T_F$.  
The free energies of stability of the {\bf NBA}, relative to the {\bf U}
states, vary within the range of $(5 - 20)k_BT$ at neutral pH [1b].
Because proteins are polymers we
expect that they would also undergo a collapse transition to a compact
phase at the temperature $T_{\theta}$ suitably modified for 
finite size systems, when water becomes a poor solvent
for the polypeptide. We have previously shown that for protein sequences
that fold in an apparent two-state manner $T_F \approx T_{\theta}$, which
naturally explains the marginal stability of proteins
\cite{Camacho93PNAS}.

The quest to understand, at the molecular level, the cooperative {\bf U}
$\leftrightarrow$ {\bf NBA} transition has lead to a number of
computational studies [2b,4,5].  
Although considerable effort has been directed to describe the molecular
basis of cooperativity, somewhat surprisingly, examination of the finite size
effects in the self-assembly of proteins has received little attention
\cite{KlimThirum02JCC}. In
contrast, scaling theories for finite sized systems undergoing regular
first and second order phase transitions have been fully developed
\cite{Fisher}.   The purpose of this paper
is to study the effect of $N$, the number of amino acid residues in a
protein, on the extent of cooperativity in the {\bf U} $\leftrightarrow$
{\bf NBA} transition.

Thermal denaturation data of wild-type (WT)  
proteins and lattice models (LMs) 
of polypeptide chains are used to examine the 
dependence of the cooperativity on 
$N$. We show that a dimensionless measure of cooperativity
\cite{KlimThirum98FD}
\begin{equation}
\Omega _c \; = \; \frac{T_F^2}{\Delta T}
|\frac{df_N}{dT}| _{T=T_F}
\label{coop_eq}
\end{equation}
grows as
\begin{equation}
\Omega _c \; \sim \; N^{\zeta}, 
\label{coop_scaling_eq}
\end{equation}
where $f_N$ is a measure of occupation of {\bf NBA}, $\Delta T$ is the full
width at half-maximum of $df_N /dT$, and $T_F$ is the folding
transition temperature identified with the maximum in  $df_N /dT$. 
We find that
\begin{equation}
\zeta \; = \; 1 + \gamma
\label{zeta_exp_eq}
\end{equation} 
where $\gamma$ is the exponent that characterizes the
divergence of susceptibility at the critical point for a $n$-component
ferromagnet with $n=0$, i.e., for a self-avoiding walk. As a byproduct of
this study we also show that $\frac{\Delta T}{T_F} \sim \frac{1}{N}$. The
parameter $\Omega_c$ is a convolution of the sharpness of the transition
($T_F/\Delta T$) and the extent to which structure, as measured by $f_N$,
changes around $T_F$. For infinite systems undergoing sharp transitions,
$\Omega _c \rightarrow \infty$, whereas $\Omega _c$ is small for broad or
highly rounded phase transitions \cite{KlimThirum98FD}. The relationship
given in Eq. (\ref{zeta_exp_eq}), which can only be valid near $T_\theta$,
establishes the proposal that $T_F\approx T_\theta$ for two-state
folders \cite{Camacho93PNAS,Thirum95}.

To establish the results given above we  used thermal and chemical
denaturation data together with  the LMs 
of a polypeptide chain to compute the growth of $\Omega _c$ with $N$. 
In the LM each amino acid is represented as a single
bead confined to the vertices of a cubic lattice [2c]. The energy of a
conformation specified by the positions, $\{\vec{r}_i\} (i = 1, 2, \ldots ,N)$,
is $E\{\vec{r}_i\}=\sum_{i<j} \, \epsilon_{ij} \delta_{r_{ij},a}$, 
where $a$ is the lattice spacing, $r_{ij} = |\vec{r}_i - \vec{r}_j|$,
$\delta_{x,a}$ is the Kronecker delta function. The contact energies
$\epsilon _{ij} = -1$, if the interaction between beads $i$ and $j$ in a
given conformation is also present in the native state (i.e., the lowest
energy conformation for a given sequence), and is zero,
otherwise. Even though simple LMs do not quantitatively 
capture the cooperativity of folding transitions in proteins 
\cite{Levitt97ProtSci}, they are 
useful for obtaining global folding properties.  The precise
choice of $\epsilon _{ij}$ should not affect the predicted universal
scaling of $\Omega _c$ with $N$.  Our purpose in undertaking LM 
Monte Carlo (MC) 
simulations is to show that Eqs.  (\ref{coop_eq})-(\ref{zeta_exp_eq})
should be valid for any model of proteins that exhibits a cooperative {\bf
U} $\leftrightarrow$ {\bf NBA} transition.

To calculate $\Omega _c$ for LMs we employ the temperature
dependence of the overlap function \cite{Camacho93PNAS}
\begin{equation}
\chi \; = \; 1 - \frac{1}{N^2-3N+2} \sum_{i<j+1}^N \,\;\delta_{r_{ij},r_{ij}^0}
\label{chi_eq}
\end{equation}
where $r_{ij}^0$ is the distance between beads $i$ and $j$ in the native
conformation. The overlap function $\chi$ 
is an order parameter that distinguishes the {\bf NBA} and {\bf U} states.  
The folding transition temperature $T_F$ can be estimated from the
location of the maximum in $d<\chi>/dT$, where $<...>$ indicates a thermal
average.  For LMs $<\chi> \approx 1- f_N$ \cite{LMCalculations}. Therefore, Eq.
(\ref{coop_eq}) may be evaluated using
\begin{equation}
\Omega _c \; = \;  \; \frac{T_F^2}{\Delta T}
\biggl(\frac{d<\chi>}{dT}\biggr)_{T=T_F}. 
\label{coop_lattice_eq}
\end{equation}

\begin{figure}
\epsfxsize=4in
\centerline{\epsffile{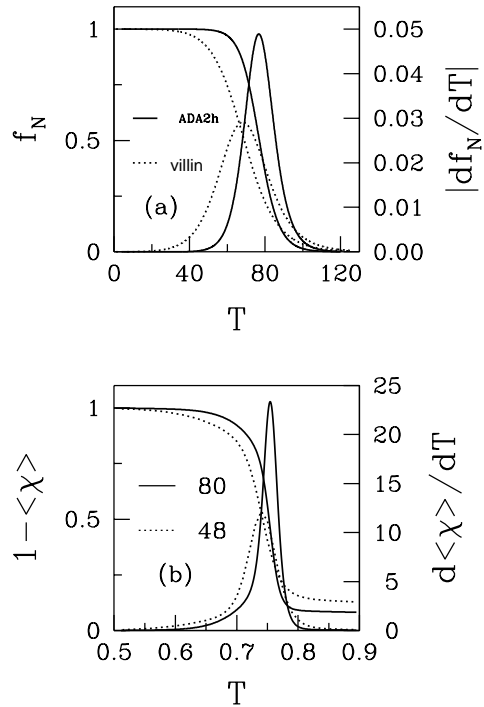}}
\caption{ (a)
Temperature dependence of the fraction of occupation of the native state,
$f_N(T)$, and its derivative $df_N/dT$.
The dotted lines are for villin head piece and the solid lines
show the data for ADA 2h \cite{ProteinDB}.
Temperature is measured in Centigrades. (b)
Dependence of 1-$<\chi>$ and $d<\chi >/dT$ on temperature
for LMs. We calculate
$\Delta T$ using $d<\chi>/dT$. The dotted lines are for the sequence with
$N=48$ and the solid lines correspond to $N=80$.
}
\label{fN_fig}
\end{figure}

{\em Analysis of experimental and simulation
data}: To establish the results given above we
first analyzed thermal denaturation data for WT proteins.  
As an example we show in Fig. \ref{fN_fig} the plot of $f_N(T)$ and
$df_N(T)/dT$ for villin ($N=35$) and ADA 2h ($N=80$)
\cite{ProteinDB}. In accord with Eq. (\ref{coop_scaling_eq}) we find that the
thermal denaturation of ADA 2h is more cooperative than that of
villin headpiece.

From Fig. 2 we find that
$\frac{\Delta T}{T_F}$, from thermal denaturation data for 32 WT
proteins \cite{ProteinDB},
scales as $N^{-\lambda}$ with $\lambda
= 1.08 \pm 0.04$.  Given that the data for
these proteins are obtained under varying experimental conditions and
using different
methods for computing the enthalpy and entropy changes at $T_F$, the
agreement between the predicted and observed behavior is excellent.
For LMs $\Delta T/T_F \sim
N^{-\lambda}$ with $\lambda=1.14\pm 0.06$ (Fig. 2). The
small deviation
of $\lambda$ from unity in LMs  is, in all likelihood, due to
the simplicity of the $\alpha$-carbon representation of the polypeptide
chain that does not capture the crucial role of side chains. 
Inclusion of side chains, which are 
tightly packed in native conformations, is expected to reduce 
fluctuations. Moreover, for $N\lesssim 40$ most of the beads are on the
surface, which also leads to considerable conformational fluctuations.  
Therefore, the expected relation $\frac{\Delta T}{T_F}
\sim N^{-1}$ holds nearly quantitatively. 

\begin{figure}
\epsfxsize=3.5in
\centerline{\epsffile{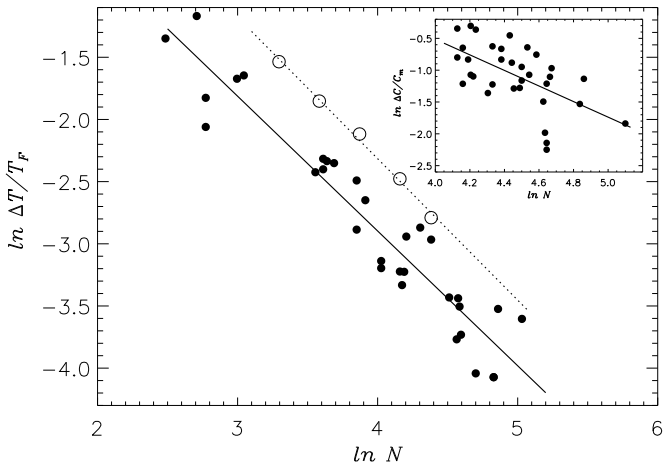}}
\caption{ The sharpness of the folding transition
$\Delta T/T_F$ as a function of $N$. Open circles represent the results from
LM simulations  with the corresponding fit (dotted line)
$\Delta T/T_F \sim N^{-\lambda}$ with $\lambda = 1.14 \pm
0.08$. The linear fit (solid line) to the experimental data for
32 WT proteins (solid circles) \cite{ProteinDB}  gives
$\lambda = 1.08 \pm 0.04$. The
correlation coefficient for  $\ln \Delta T/T_F$ and $\ln N$ is 0.95.
For clarity
LM  data are shifted up by 0.4. Inset shows the dependence of the
width of folding transition $\ln \Delta C/C_m$
for chemical denaturation on $\ln N$. The
linear fit to the data points collected for 33 WT  proteins
yields $\lambda = 1.22\pm 0.14$ (the correlation factor is 0.59).
}
\end{figure}

The dependence of $\Omega_c$ on $N$ for WT  proteins and LMs shows
that $\Omega_c \sim N^{\zeta}$ (Fig. (3)).   From the linear fit to
the log-log plot of the data  we find $\zeta \approx 2.17\pm 0.09$ for
WT proteins and $\approx 2.33\pm 0.08$ for LMs. The 5th order
$\epsilon$ expansion for polymers using $n$-component $\phi^4$ theory
with $n=0$ gives $\gamma=1.22$ \cite{Kleinert}. Thus, from
Eq. (\ref{zeta_exp_eq}) we predict that $\zeta \approx 2.22$.  Thus,
the data for  WT proteins and LMs  are consistent with the
theoretical prediction (Eq. (\ref{zeta_exp_eq})).  We should emphasize
that the robustness of the fit has been checked using different
fitting procedures.  The {\em remarkable finding relating the critical
exponent $\gamma$ to thermal denaturation of proteins} gives further
credence to the proposal that efficient folding is achieved at $T_F
\approx T_{\theta}$ \cite{Camacho93PNAS}. 
It also suggests that {\bf U} $ \leftrightarrow$
{\bf NBA} transition is only weakly first order, thus explaining the
marginal stability of globular proteins.

\begin{figure}
\epsfxsize=3.5in
\centerline{\epsffile{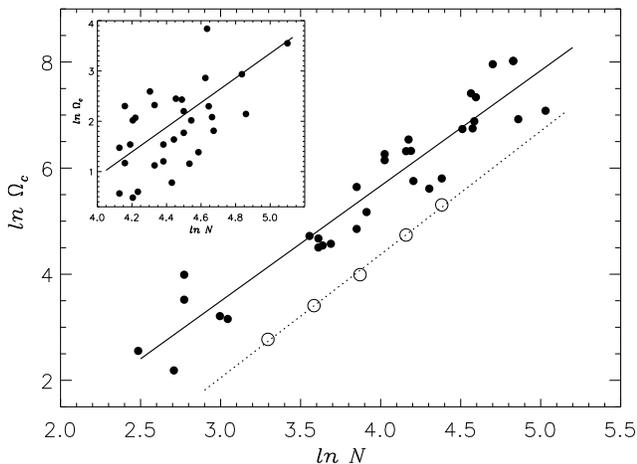}}
\caption{ Plot of ln$\Omega_c$ as a function of ln$N$.
Symbols are the same as in Fig. (2).
The dotted line is a  fit to the LM data, which gives
$\zeta =2.33 \pm 0.08$.
The solid line is a fit to the experimental values of
$\Omega_c$ \cite{ProteinDB} with the  exponent
$\zeta = 2.17 \pm 0.09$. The correlation coefficient for
$\ln \Omega_c$  and $\ln N$ is 0.95.
The LM data are shifted down by 0.7. The dependence of the folding
cooperativity  on $N$ for chemical denaturation is plotted in the
inset. The linear fit to experimental data (solid line) results in
the exponent $\zeta = 2.45\pm 0.29$ (the correlation factor is 0.59).
Both sets of experimental data rule out $\zeta=2$.
}
\end{figure}

Most folding experiments are performed by titrating with
denaturants (urea or guanidine hydrochloride). At denaturant
concentrations above the midpoint $C_m$ (at which the populations of the
folded and unfolded states are equal) proteins are denaturated. Thus,
phase transitions to the {\bf NBA} occur by varying denaturant
concentration. In analogy with Eq. (\ref{coop_eq}) we computed, for
33 WT proteins \cite{ProteinDB}, $\Omega _c = \frac{C_m^2}{\Delta C}
| \frac{df_N}{dC}| _{C=C_m}$ and $\frac{\Delta C}{C_m}$, where
$\Delta C$ is the full width at half-maximum of $df_N/dC$. The plots of
ln$\frac{\Delta C}{C_m}$ and ln$\Omega_c$ as a function of ln$N$ yield
$\lambda \approx 1.22\pm0.14$ and  
$\zeta \approx  2.45\pm0.29$, respectively (see the
insets to Figs. (2,3)).  Thus, the
scaling of $\Omega_c$ and $\Delta C/C_m$ remains essentially
unchanged even {\em  though the
chemical and thermal denaturation mechanisms are vastly different}. This
result also suggests  that $\zeta$ is universal. However, the dependence
of ln$\Omega_c$ on ln$N$ has a correlation coefficient of only about 0.6
compared to 0.95 for thermal denaturation. We believe that larger
uncertainties result from greater drift in the
experimental signals in denaturant-induced unfolding compared to thermal
denaturation \cite{Camacho93PNAS}.

The rationale for Eqs. (2,3) 
is based on the following arguments. (1) By analogy with
magnetic systems $\Delta \chi$ is similar to susceptibility and should be
given by $\Delta \chi = T\partial <\chi>/\partial h$, where $h$ is a
"magnetic" or an ordering field conjugate to $\chi$.  Because $\Delta
\chi$ is dimensionless, we expect that the ordering field $h \sim T$ and
thus $Td<\chi>/dT$ in proteins is similar to magnetic susceptibility. (2)
Camacho and Thirumalai  \cite{Camacho93PNAS} have suggested that
efficient folding in apparent two-state folders requires $T_F\approx
T_\theta$. Because the transition at $T_\theta$ is usually second order
\cite{GrosbergBook}, while the one at $T_F$ is first order [2c,11],
the $T_F\approx T_\theta$ condition implies that folding of two-state
globular proteins occurs near a tricritical point \cite{Camacho93PNAS}.  
Therefore, the critical exponents that control the
behavior of the polypeptide chain at $T_{\theta}$ should manifest itself
in the {\bf U} $\leftrightarrow$ {\bf NBA} phase transition. Using these
arguments we can obtain the $N$ dependence of $\Omega_c$ in the following
way. In general, we expect that close to $T \approx T_{\theta} \approx
T_F$ the Flory radius \cite{DeGennesBook} $R_F \sim \Delta T^{-\nu} \sim
N^{\nu}$ ($R_F$ is the analogue of the correlation length in magnetic
systems). This implies that $\Delta T/T_F \sim N^{-1}$. Because of the
analogy to magnetic susceptibility, we expect $Td<\chi>/dT \sim N^\gamma$.
Using Eq. (\ref{coop_lattice_eq}) we obtain the expected relationship
$\Omega_c \sim N^{1+\gamma}$, which directly follows from the 
hypothesis that $T_F\approx T_\theta$ for efficient two-state folders
\cite{Camacho93PNAS}.

The scaling $\Omega_c \sim N^{\zeta}$ with $\zeta$ clearly different
from 2 may appear to be at odds with the idea that the structures in 
the NBA are sequence-specific. However, the global characteristics embodied
in the growth of $\Omega_c$ with $N$ are {\em valid only at} $T\approx T_F$. In
the neighborhood of this temperature the general characteristics
of the {\bf U} $\leftrightarrow$ {\bf NBA} transition are governed
by the properties of the unfolded states as $T_F$ is approached from above.
It has been shown that in the denaturated states ($T>T_F$) the global
properties like the gyration radius $R_g \sim N^{\nu}$ with
$\nu \approx 0.59$ {\em as expected for homopolymers} \cite{Kohn04}.
Similarly, the homopolymeric 
nature around $T_F$ is reflected in the growth of $\Omega_c$ with $N$. 

The finding that the folding transition
at $T_F$ occurs at a tricritical point suggests
that the native states of natural proteins are only marginally stable.
Because biological functions require transitions between different states,
it is logical to postulate that natural foldable proteins have evolved to
ensure $T_\theta \approx T_F$. The coil-globular transition at
$T_\theta$ is likely to be a second order transition involving no
discontinuity in the free energy. At $T_F$ the transition is of the
first order. The closeness of $T_F$ and $T_\theta$ implies that the
discontinuity of the free energy at $T_F$ cannot be large. As a result
the folded state is expected to be only marginally stable with respect
to the ensemble of denatured states. As argued elsewhere
\cite{Thirum95} this
condition is also equivalent to maximizing the ratio $T_F/T_g$, where
$T_g$ is a glass transition temperature \cite{Onuchic97}. 
Marginality condition may also
be a requirement for robustness of the folded state. This may explain why
small single domain proteins can tolerate a large number of mutations
without substantial changes in the native state.
It is also likely, as recently shown, that evolution has led to marginally
stable proteins that have maximum sequence-structure 
compatibility \cite{Xia02,Taverna02}.

This work was
supported in part by a KBN grant  No 1P03B01827
 and the National Science Foundation grant
(NSF CHE-0209340). We are grateful to R.B. Dyer, A.R. Fersht, and
N. Ferguson for providing us with unpublished thermal denaturation data. 
Mai Suan Li wishes to thank the hospitality of
IPST, Maryland, USA and ICTP, Trieste, where part of this work was done.


\begin{references}
\bibitem{LMCalculations}
We generated between 12 to 20 distinct maximally compact native
structures for a given value of $N$ ($27\le N\le 80$).  For example,
all native structures for $N=80$ were confined to  the vertices of a
4x4x5 cube. In the Go model each structure corresponds to a
distinct sequence. We used, depending on $N$, between 50 and 100 Monte
Carlo trajectories to collect states and applied multiple histogram method
\cite{Ferrenberg} to compute $<\chi(T)>$,  $d<\chi>/dT$,  and 
$\Omega_c$ using Eq. (5). For the Go models (and other 
highly optimized sequences) $T_F$, identified with the location of the
maximum in $d<\chi>/dT$, coincides with $T_\theta$
\cite{Camacho93PNAS}.
\end{references}
\end{document}